\documentclass[%
 reprint,
 amsmath,amssymb,
 aps,
]{revtex4-2}

\usepackage{epsfig,amssymb,amsmath,graphicx,subfigure,hyperref}
\usepackage{tabularx}
\usepackage{float}
\usepackage{setspace}
\usepackage{color}
\usepackage{array} 
\newcommand{\PreserveBackslash}[1]{\let\temp=\\#1\let\\=\temp} \newcolumntype{C}[1]{>{\PreserveBackslash\centering}p{#1}} \newcolumntype{R}[1]{>{\PreserveBackslash\raggedleft}p{#1}} \newcolumntype{L}[1]{>{\PreserveBackslash\raggedright}p{#1}} 
\usepackage{graphicx}
\usepackage{dcolumn}
\usepackage{bm}
\usepackage{comment}
\usepackage{color}
\usepackage{soul}

\begin{document}

\preprint{APS/123-QED}

\title{Self-propelled particles undergoing cyclic transitions}


\author{Ye Zhang}
\affiliation{Key Laboratory of Artificial Micro- and Nano-structures of Ministry of Education and School of Physics and Technology, Wuhan University, Wuhan 430072, China}
\author{Duanduan Wan}
\email[E-mail: ]{ddwan@whu.edu.cn}
\affiliation{Key Laboratory of Artificial Micro- and Nano-structures of Ministry of Education and School of Physics and Technology, Wuhan University, Wuhan 430072, China}

\date{\today}

\begin{abstract}

Cyclic transitions between active and passive states are central to many natural and synthetic systems, ranging from light-driven active particles to animal migrations. Here, we investigate a minimal model of self-propelled Brownian particles undergoing cyclic transitions across three spatial zones: gain, loss, and neutral regions. Particles become active in the gain region, passive in the loss region, and retain their state in the neutral region.
By analyzing the steady-state behavior as a function of particle number and the size of the loss region, we identify a threshold particle number, below and above which distinct structural changes are observed. Interestingly, below this threshold, increasing the particle number reduces the state-switching time (the time required for a particle to transition from active to passive and back to active). In contrast, above the threshold, further increases in particle number result in longer switching times. In the subthreshold regime, our analytical model predicts structural characteristics and switching dynamics that align well with simulations. Above the threshold, we observe an emergent spatial clustering, with particles transitioning from passive to active states in close proximity. These findings provide insights into the collective dynamics of cyclic processes between active and passive states across distinct spatial zones in active matter systems.

\end{abstract}

\maketitle



Self-propelled particle models are widely used to describe the collective dynamics of active matter systems, from microscopic  motors \cite{ Cisneros_2007_Fluid,Guix_2018_Self-Propelled} and bacterial colonies \cite{Chen_2017_Weak, Liu_2021_Viscoelastic, Zhang_2010_Collective} to macroscopic phenomena such as bird flocks and animal herds \cite{Vicsek_1995_Novel, Gregoire_2004_Onset, Marchetti_2013_Hydrodynamics, Gompper_2024_2024}. These systems exhibit a broad range of emergent behaviors, including flocking \cite{Vicsek_1995_Novel, Liebchen_2017_Collective, Caussin_2014_Emergent}, motility-induced phase separation (MIPS) \cite{Buttinoni_2013_Dynamical, Redner_2013_Structure, Elgeti_2015_Physics, Bechinger_2016_Activeb, Cates_2015_Motility-Induced, Digregorio_2018_Full, Omar_2021_Phasea}, and glassy states \cite{Redner_2013_Structure, Berthier_2013_Non-equilibrium, Szamel_2015_Glassy, Wiese_2023_Fluid-Glass-Jamming}. Confinement and boundary interactions \cite{Elgeti_2013_Wall, Fily_2014_Dynamicsa, Williams_2022_Confinementinduced, Liu_2016_Bimetallic}, mixtures of active and passive particles \cite{Stenhammar_2015_ActivityInduced, Williams_2022_Confinementinduced}, and spatially varying activity \cite{Palacci_2013_Living, Frangipane_2018_Dynamica, Farrell_2012_Patterna} further contribute to the dynamical complexity of these systems.

Cyclic transitions between active and passive states are observed in many active matter systems. For instance, gain-loss cycles occur in light-driven active particles \cite{Rey_2023_Light, Frangipane_2018_Dynamica}, where particles gain motility in illuminated regions, lose it in shaded regions, and regain it upon returning to the illuminated zones. Similarly, dung beetles exhibit cyclical behavior, moving quickly to a manure pile and returning more slowly while pushing a dung ball \cite{Dacke_2020_dunga}. Bees also demonstrate a comparable cycle, rapidly flying between flower-rich zones to collect nectar, then returning more slowly to their hives for subsequent foraging. Migratory birds alternate between gaining energy at resource-rich stopovers, expending it during long-distance flights, and replenishing their reserves upon revisiting stopovers \cite{Lincoln_1979_Migration}. These examples highlight the cyclic nature of transitions between active and passive states across distinct spatial regions. While much work has focused on static gradients or spatially varying activity \cite{Rey_2023_Light, Frangipane_2018_Dynamica, Varuni_2017_Phototaxis, Ghosh_2015_Pseudochemotactic, Luchsinger_2000_Transport}, the study of cyclic transitions between active and passive states across distinct spatial regions remains scarcely explored.

\begin{figure}
\centering 
\hspace{.0cm}
\includegraphics[width=3.3 in]{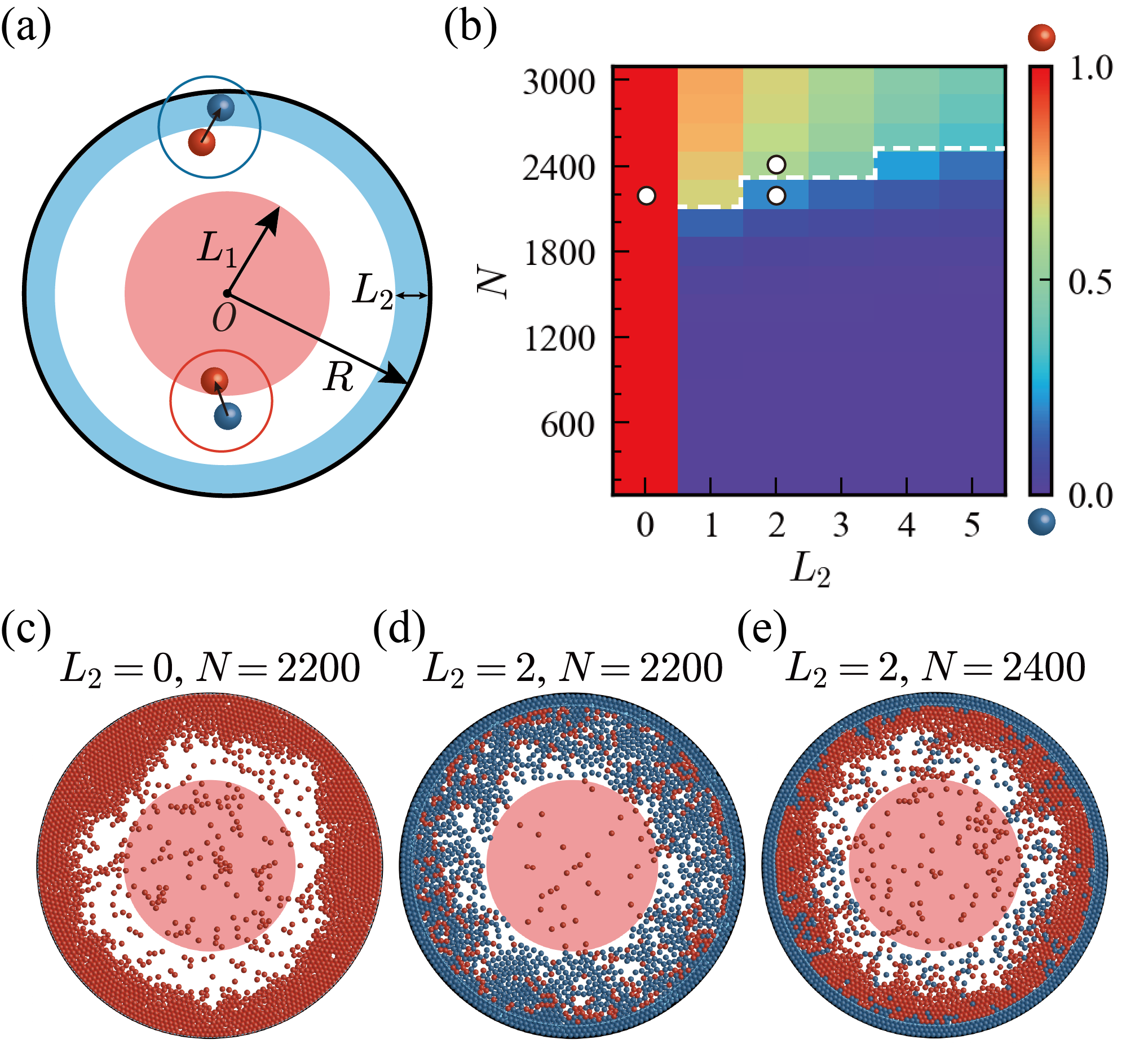}
\caption{ (a) Illustration of the system. The black boundary indicates the hard wall. Pink, light blue, and white regions represent the gain, loss, and neutral areas, respectively. Red and blue spheres represent particles in the active and passive states, respectively. When a particle enters the active region, it becomes active (red circle); when it enters the passive region, it becomes passive (blue circle). (b) Fraction of particles in the active state as a function of $L_2$ and the total number of particles $N$. Snapshots are taken every $0.01\tau$ after the system reaches steady state, with data averaged over $10^4$ snapshots. The white dashed line indicates the estimated transition point. The three white circles highlight three selected data points. (c-d) Snapshots of the steady states corresponding to the three white circles in (b).}
\label{fig_illustration}
\end{figure}

In these systems, the duration of the cycle—referred to as state-switching time—plays a crucial role in determining overall efficiency and stability. Longer cycle times may introduce inefficiencies or energy loss, while shorter cycle times could lead to overly rapid transitions that destabilize the system. Despite the importance of these cycles in natural and synthetic systems, the state-switching time has not been  explored in the context of cyclic transitions. Investigating these cycles offers valuable insights into the optimization and regulation of dynamic systems and provides a fresh perspective on how cyclic behavior governs the performance of self-propelled systems.

In this study, we investigate a minimal model of self-propelled Brownian particles undergoing cyclic transitions between active and passive states across distinct regions. Particles become active in the gain region, passive in the loss region, and remain in their state in the neutral region. By analyzing the system in terms of particle number and the size of the loss region, we identify a threshold particle number that separates the system into two distinct regimes. Below this threshold, increasing the particle number reduces the state-switching time (i.e., the time required for a particle to transition from active to passive and back to active). However, above the threshold, further increases in particle number result in longer switching times. Additionally, above the threshold, we observe the emergence of spatial clustering, where particles transitioning from passive to active states tend to do so in close proximity. These findings reveal rich dynamical features arising from cyclic activity processes.


To investigate these cyclic behaviors, we model a system of $N$ Brownian particles confined within a circular region of radius $R$, surrounded by a hard wall, as shown in Fig.~\ref{fig_illustration}(a). This region is divided into three distinct zones: gain, loss, and neutral, represented by pink, light blue, and white areas, respectively. The gain zone has a radius $L_1$, and the loss zone has a width $L_2$. The dynamics of the particles are governed by the overdamped Langevin equations:
\begin{equation}
 \mathbf{\dot{r} }_i=D_\text{t}\beta \{f_i \mathbf{n}_i-\nabla _i[\sum_{j(\neq i)}U(r_{ij}) + U(r_{iw}) ] 
 \} +\sqrt{2 D_\text{t}}\bm{\eta}^t_i , 
\label{eq:theta} 
\end{equation}
\begin{equation}
\dot{\theta}_i=\sqrt{2 D_\text{r}}\eta^r_i .
\end{equation}
Here $\mathbf{r}_i$ and $\mathbf{n}_i = (\cos \theta_i, \sin \theta_i) $ denote the position and self-propulsion direction of the $i$-th particle, respectively. $\beta= 1/k_B T$, with $k_B$ being the Boltzmann constant and $T$ the temperature. The Weeks–Chandler–Anderson potential $U(r)$ models excluded-volume repulsive interactions between particles and between particles and the hard wall:
\begin{equation}
 U(r)=4\epsilon[(\frac{\sigma}{r} )^{12}-(\frac{\sigma}{r})^6]+\epsilon,  \, r< 2^{1/6}\sigma,
\end{equation}
with $r_{ij}=|\mathbf{r}_i-\mathbf{r}_j|$ and $r_{iw} = R + 2^{1/6}\sigma/2 - |\mathbf{r}_i|$. We set $\epsilon = k_B T$. The translational noise $\bm{\eta}^t$ and rotational noise $\eta^r$ have zero mean and unit variance, with $\langle \bm{\eta}^\mathrm{t}_i(t) \rangle = \mathbf{0}$, $\langle  \bm{\eta}^\mathrm{t}_i(t)\bm{\eta}^\mathrm{t}_j(t')\rangle = \delta_{ij}\delta(t-t') \mathbf{1}$, $\langle  \eta_i^\mathrm{r} (t) \rangle = 0$, and $\langle  \eta_i^\mathrm{r}(t)\eta_j^\mathrm{r}(t') \rangle = \delta_{ij}\delta(t-t')$.
The translational diffusion coefficient $D_\text{t}$ and rotational diffusion coefficient $D_\text{r}$ are related by $D_\text{t} = D_\text{r} \sigma^2 /3$. We adopt $\sigma$ and $k_B T$ as our units of length and energy, respectively, and $\tau=\sigma^2/D_\text{t}$ as the unit of time, with $D_\text{t} = 1$. 
Particles adopt $f_i = f_{\text{0}}$ upon entering the gain area and $f_i = 0$ upon entering the loss area, maintaining these values while inside and after exiting the respective regions. In the simulations, we set $R = 30$, $L_1 = 15$, $f_{\text{0}} = 150$, and vary $N$ and $L_2$. Particles are initially randomly distributed with random orientations. Numerical integration is performed using HOOMD-blue’s overdamped Brownian module \cite{Anderson_2020_HOOMDblue} with a time step of $\text{d}t = 10^{-6} \tau$. Simulations run for at least $200\tau$.

We present the steady-state fraction of active particles as a function of loss area width $L_2$ and particle number $N$ in Fig.~\ref{fig_illustration}(b), with $L_2$ ranging from 0 to 5 and $N$ from 200 to 3000. Three representative steady-state configurations, highlighted by the white circles in Fig.~\ref{fig_illustration}(b), are shown in Figs.~\ref{fig_illustration}(c-e). As seen in Fig.~\ref{fig_illustration}(b), when $L_2 = 0$ (no loss region), all particles remain active, as any particle entering the gain region stays active indefinitely. A corresponding steady-state snapshot, shown in Fig.~\ref{fig_illustration}(c), illustrates active particles accumulating at the boundary, consistent with previous observations \cite{Yang_2014_Aggregation}.
For nonzero $L_2$, the fraction of active particles exhibits a transition as $N$ increases. Below the transition, the active fraction is low, while above it, there is a sharp increase. This transition is indicated by the white dashed line in Fig.~\ref{fig_illustration}(b). At $N$ values well below the transition, most particles are passive and scattered across the loss and neutral regions, with only a few active particles in the gain region (see Supplemental Materials). As $N$ approaches the transition, the active fraction increases in the neutral region near the loss boundary [Fig.~\ref{fig_illustration}(d)]. Beyond the transition, the number of active particles increases significantly, accumulating near the boundary of the loss region [Fig.~\ref{fig_illustration}(e)]. For larger $N$, multiple layers of active particles form (see Supplemental Materials).

\begin{figure}
\centering 
\includegraphics[width= 3.4 in]{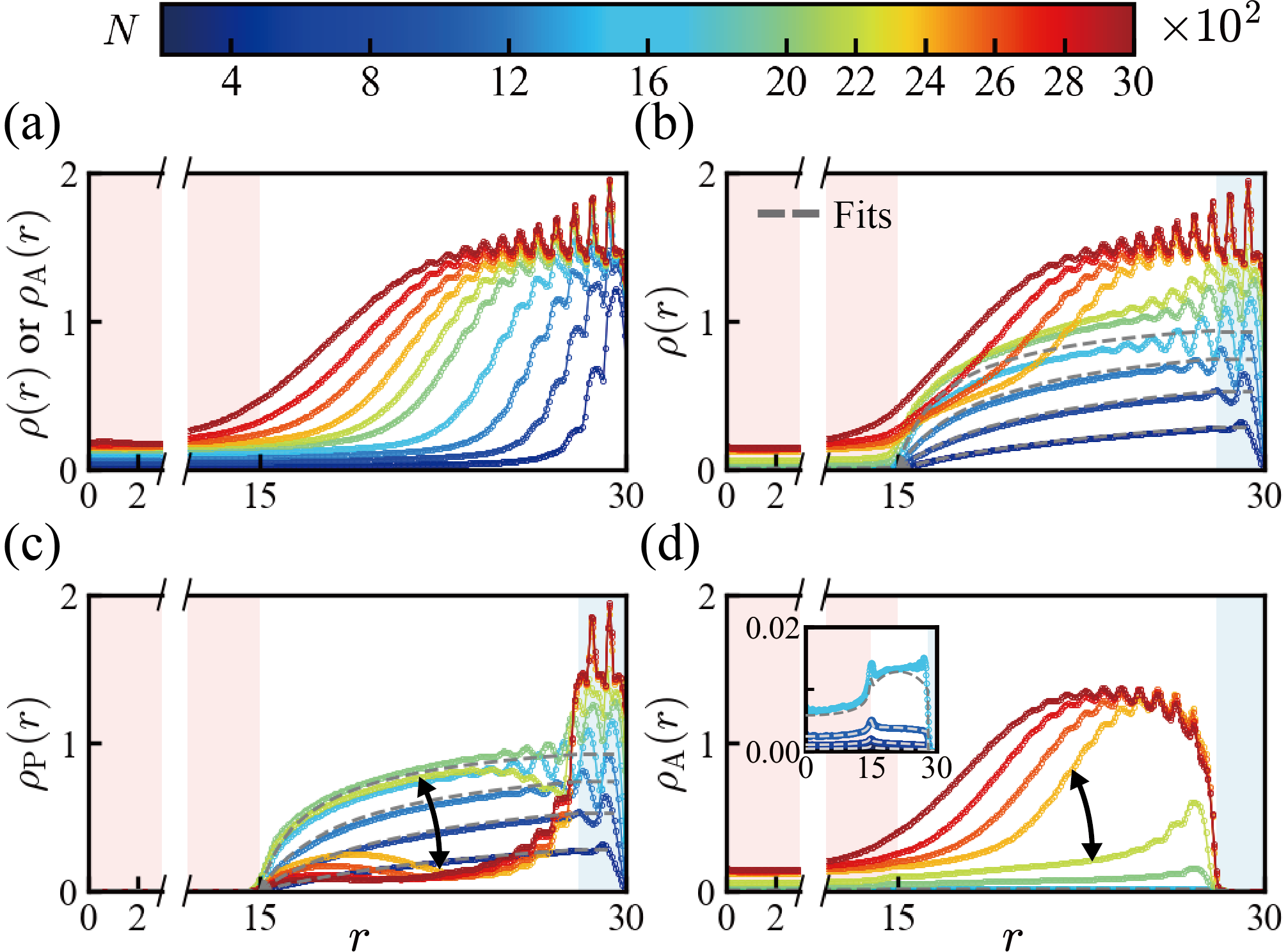}
\caption{\label{fig:density} Particle number densities in the radial direction. (a) Total particle density for $L_2 = 0$. (b-d) For $L_2 = 2$, showing the total particle density (b), and the densities of particles in the passive (c) and active (d) states. The pink and light blue shaded regions indicate the gain and loss zones, respectively. Symbols represent simulation results, with solid lines for clarity. Dashed lines correspond to model predictions. Black arrows in (c) and (d) indicate the transition. The inset in (d) provides an enlarged view of the low-density curves for improved visibility.}
\label{fig_density}
\end{figure}

To investigate the structural characteristics of the system's steady states, Fig.~\ref{fig_density} presents the particle number density in the radial direction, calculated from the time-averaged number of particles in a given area (see Supplemental Material for details). Figure \ref{fig_density}(a) shows the case for $L_2 = 0$, where no loss region exists. In this case, all particles become active, so the total density $\rho(r)$ is identical to the density of active particles $\rho_A(r)$. As shown in Fig.~\ref{fig_density}(a), $\rho(r)$ exhibits peaks near the boundary wall, with peak amplitudes decaying as the distance from the wall increases. As $N$ increases, the peaks become sharper, and the density oscillations extend further into the system.
Figures \ref{fig_density}(b-d) show the case for $L_2 = 2$, where a loss region is present. Panel (b) shows $\rho(r)$, the total particle density, while panels (c) and (d) show the densities of passive and active particles, $\rho_P(r)$ and $\rho_A(r)$, respectively. For $L_2 = 2$, density oscillations again begin near the wall, as seen in the $\rho(r)$ plot. A significant change occurs as $N$ increases from 2200 to 2400 (see the yellow and light green lines), crossing the transition. For $N \leq 2200$, the number of peaks near the wall increases from one to several as $N$ grows from 400. The peaks remain broad, and the density decays slowly beyond the oscillations. For $N \geq 2400$, the peaks become sharper and extend farther from the wall, with the density decaying rapidly beyond the peaks.
This transition is particularly evident in the $\rho_P(r)$ and $\rho_A(r)$ profiles. In the neutral region [white background in Figs.~\ref{fig_density}(b-d), $15 \leq r \leq 28$], adjacent to the loss region, $\rho_P(r)$ decreases significantly as $N$ increases from 2200 to 2400 [as indicated by the black arrow in Fig.~\ref{fig_density}(c)], while $\rho_A(r)$ shows a sharp increase [as indicated by the black arrow in Fig.~\ref{fig_density}(d)], marking the transition.

\begin{figure}[tbp]
\centering 
\hspace{-0.5cm}
\includegraphics[width=3.2 in]{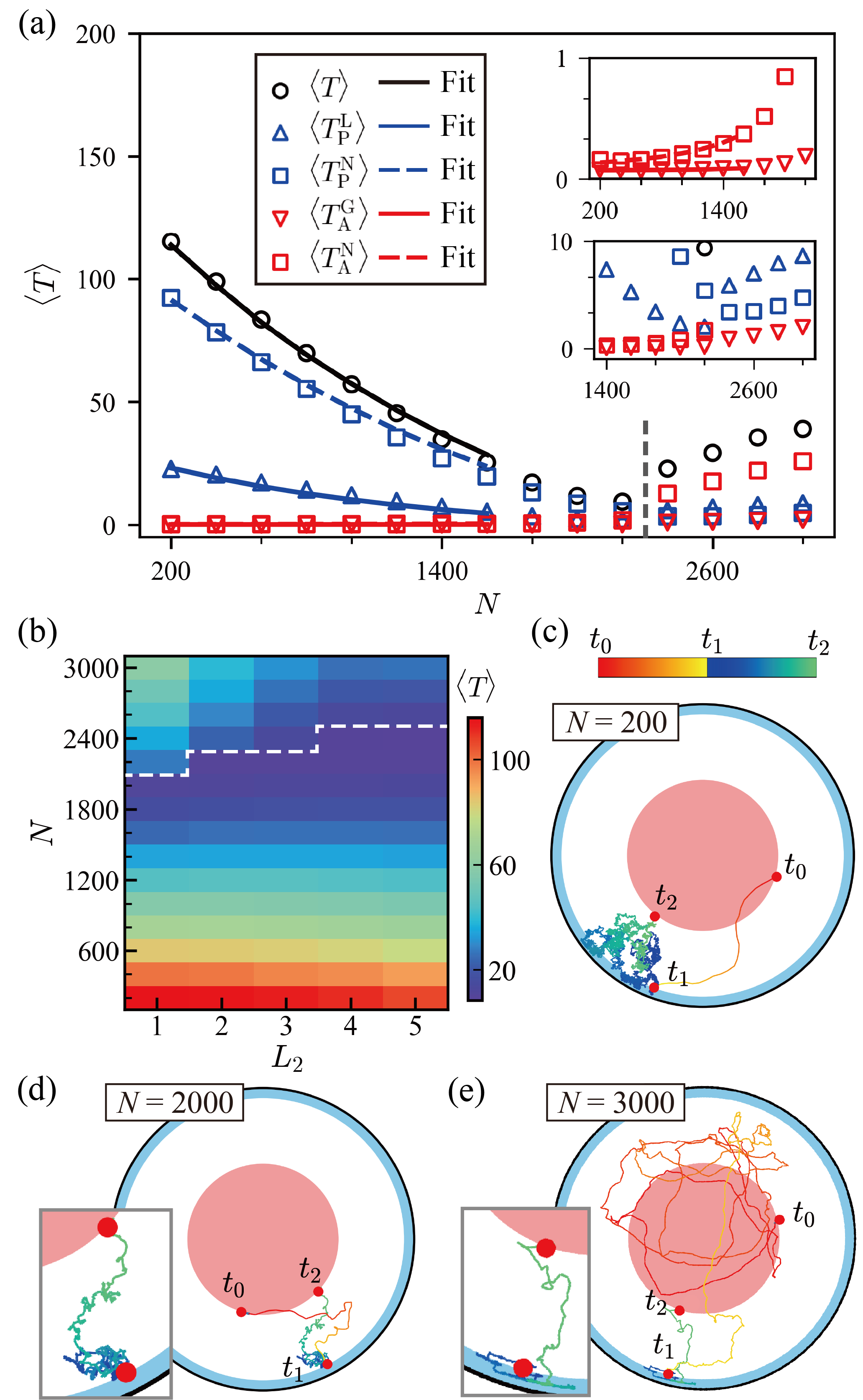}
\caption{(a) For $L_2 = 2$, the quantities $\langle T \rangle$, $\langle T_{\text{P}}^{\text{L}} \rangle$, $\langle T_{\text{P}}^{\text{N}} \rangle$, $\langle T_{\text{A}}^{\text{G}} \rangle$, and $\langle T_{\text{A}}^{\text{N}} \rangle$ are plotted as functions of $N$. Symbols represent simulation results, with fitting lines corresponding to model predictions. The vertical grey dashed line marks the transition. The vertical grey dashed line marks the transition point. Insets show enlarged views for clarity. (b) $\langle T \rangle$ plotted as a function of $L_2$ and $N$. For each data point, at least $10^4$ cycles are analyzed to estimate $\langle T \rangle$. (c-e) Trajectory of a particle over a period, starting from an active state at time $t_0$, transitioning to passive at time $t_1$, remaining passive, and then returning to the active state at time $t_2$ when the recording concludes. Insets in (d) and (e) provide enlarged views of the trajectory from $t_1$ to $t_2$ for better visibility.}
\label{fig_duration}
\end{figure}

Next, we explore the dynamics of the system. In steady states with nonzero $L_2$, particles exhibit cyclic motion, alternating between active and passive states. To characterize this behavior, we define the average state-switching time, $\langle T \rangle$, as the average duration for a particle to complete a full cycle: transitioning from active to passive and back to active. We also define four components of this cycle: $\langle T_{\text{P}}^{\text{L}} \rangle$, $\langle T_{\text{P}}^{\text{N}} \rangle$, $\langle T_{\text{A}}^{\text{G}} \rangle$, and $\langle T_{\text{A}}^{\text{N}} \rangle$, representing the average times spent by passive particles in the loss region, passive particles in the neutral region, active particles in the gain region, and active particles in the neutral region, respectively.
For $L_2 = 2$, we plot $\langle T \rangle$ and its components as functions of $N$ in Fig.~\ref{fig_duration}(a). Interestingly, below the transition (indicated by the vertical grey dashed line), $\langle T \rangle$ decreases with increasing $N$, while above the transition, $\langle T \rangle$ increases with $N$ [see black circles in Fig.~\ref{fig_duration}(a)]. This trend holds for other $L_2$ values as well. Figure \ref{fig_duration}(b) presents $\langle T \rangle$ as a function of $L_2$ and $N$. The dashed white line indicates the transition point, below which $\langle T \rangle$ increases with $N$, and above which $\langle T \rangle$ decreases with $N$. Notably, this transition line coincides with the position identified in Fig.~\ref{fig_illustration}(b).

To further understand the dynamic motion of particles during state-switching, we plot the trajectories of individual particles in Fig.~\ref{fig_duration}(c-e). As shown in Fig.~\ref{fig_duration}(c), when $N$ is small, the trajectory from the particle's initial active state to becoming passive (from $t_0$ to $t_1$) is relatively smooth and direct. After entering the loss region and becoming passive at $t_1$, the trajectory from $t_1$ to $t_2$—when the particle re-enters the gain region and reactivates—becomes more winding, resembling random Brownian-like motion.
As $N$ increases but remains below the transition, the trajectory from $t_0$ to $t_1$ exhibits undulations, while the trajectory from $t_1$ to $t_2$ becomes more linear and less convoluted, reflecting reduced random wandering [Fig.~\ref{fig_duration}(d)]. After the transition, the trajectory from $t_0$ to $t_1$ spans a larger region of the system, showing more extensive windings, while the $t_1$ to $t_2$ trajectory becomes straighter and more streamlined [Fig.~\ref{fig_duration}(e)].

\begin{figure}[] 
\centering 
\includegraphics[width=3.3 in]{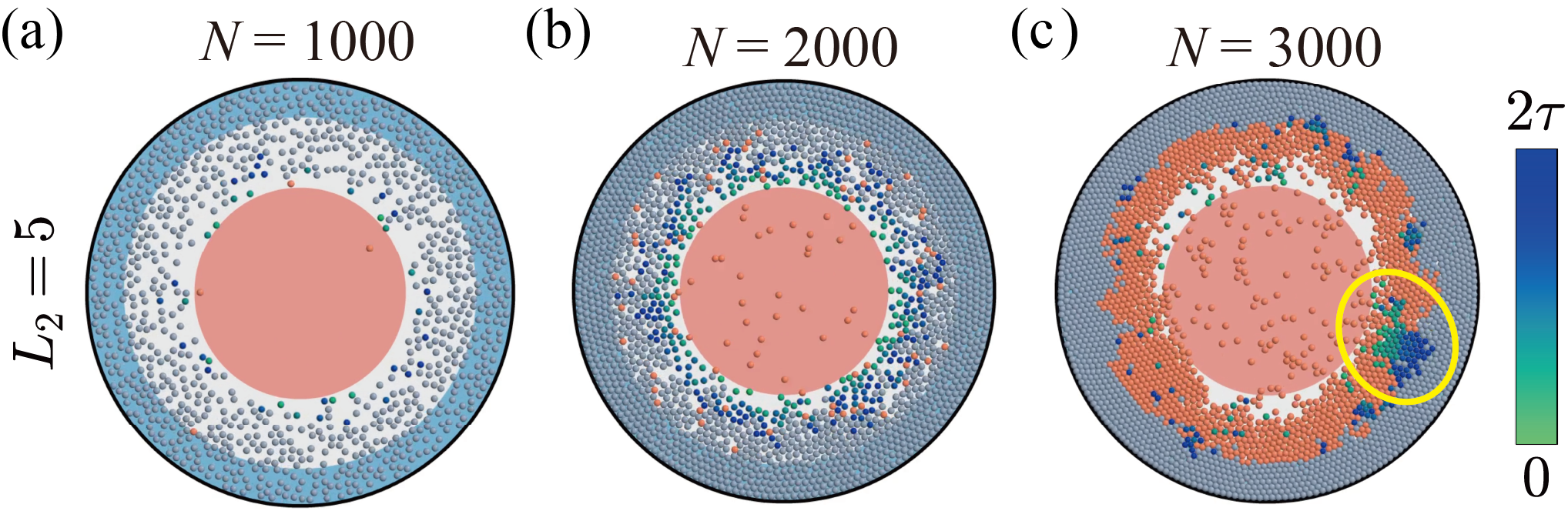}
\caption{The sequence of particles transitioning from passive to active states. Active particles are marked in pink, while passive particles that will become active within the next $2\tau$ are color-coded using a gradient: green particles will activate soon, while blue ones will activate closer to $2\tau$. Other passive particles are shown in grey. Yellow circle highlights the clustering phenomenon, where particles about to become active are in close proximity.}
\label{fig_clustering}
\end{figure}

We now proceed with an analytical analysis of the system, focusing on the regime where the particle number is below the transition threshold. Previous studies \cite{Bruna_2012_Excluded-volume, Bruna_2012_Diffusion} show that excluded-volume effects influence diffusion in hard spheres, leading to a nonlinear diffusion equation for the one-particle distribution function with an enhanced collective diffusion rate. In our case, due to the relatively large self-propulsion velocity of active particles and the steady-state condition below the transition (where active particles comprise only a small fraction of the total), we neglect their diffusion terms. Instead, we focus on an effective drift term arising from collisions with passive particles.
Thus, the diffusion equation for the probability density function $\psi_{\text{A}, i}(x, y, \theta, t)$ of an active particle $i$ is given by:
\begin{equation} 
\frac{\partial \psi_{\text{A},i}}{\partial t} = \nabla \cdot \left( -v_{\text{eff}} \mathbf{n}_i \psi_{\text{A},i} \right) + D_{\text{r}} \frac{\partial^2 \psi_{\text{A},i}}{\partial \theta^2},
\label{eq_active} 
\end{equation}
where $v_{\text{eff}}$ is the effective self-propulsion velocity of an active particle, which deviates from the original self-propulsion velocity $v_0 = D_\text{t}\beta f_0$ due to collisions with passive particles.
We determine $v_{\text{eff}}$ by measuring the instantaneous velocity projected onto the self-propulsion direction: 
\begin{equation} 
v_{\text{eff}} = \langle \dot{\mathbf{r}}_i \cdot \mathbf{n}_i \rangle = \langle (\mathbf{d}_i / \text{d}t) \cdot \mathbf{n}_i \rangle. 
\label{Eq:veff} 
\end{equation} 
Here $\mathbf{d}_i$ denotes the displacement of particle $i$ over a time step $\text{d}t$. Simulation results indicate that $v_{\text{eff}}$ decreases linearly with increasing passive particle density $\rho_{\text{P}}$, consistent with previous studies \cite{Stenhammar_2013_Continuum, Stenhammar_2015_ActivityInduced}. Fitting the simulation data to the relation $v_{\text{eff}} = (1 - c \rho_{\text{P}}) v_0$ with $v_0 = 150$ yields $c = 0.75$. Collisions with passive particles reduce the active particle's self-propulsion by an obstructive force $c \rho_{\text{P}} v_0 / D_\text{t}\beta$, while each passive particle experiences an equal reaction force $c v_0 / D_\text{t}\beta$ in the active particle's direction. Thus, the diffusion equation  for the probability density function of a passive-state particle, $\psi_{\text{P},i}$, is given as :

\begin{equation}
\begin{aligned}
\frac{\partial \psi_{\text{P},i}}{\partial t} = & \nabla \cdot \left( D_{\text{c}}(\rho_{\text{P}}) \nabla \psi_{\text{P},i} - \sum_{j} \int_{0}^{2\pi} \psi_{\text{A}, j} c v_0 \mathbf{n}_j \text{d}\theta \psi_{\text{P},i}   \right) \\
& + D_{\text{r}} \frac{\partial^2 \psi_{\text{P},i}}{\partial \theta^2}.
\label{eq_passive} 
\end{aligned}
\end{equation}
The second term on the right-hand side of Eq.~(\ref{eq_passive}) represents the effective force exerted by active particles on a passive particle. The collective diffusion coefficient, $D_{\text{c}}(\rho_{\text{P}})$, depends on the passive particle density $\rho_{\text{P}}$. We compute $D_{\text{c}}(\rho_{\text{P}})$ using the static structure factor and the isothermal compressibility $\kappa_{\text{T}}$ \cite{Hess_1983_Generalized, Lahtinen_2001_Diffusion}, with $\kappa_{\text{T}}$ derived from the equation of state for hard disks \cite{Douglas_1975_ASimple}. The expression for $D_{\text{c}}(\rho_{\text{P}})$ is:
\begin{equation*} 
D_{\text{c}}(\rho_{\text{P}}) = \frac{D_{\text{t}} \beta}{\rho_{\text{P}} \kappa_{\text{T}}} = \frac{\pi^3 \rho_{\text{P}}^3 - 12 \pi^2 \rho_{\text{P}}^2 - 128 \pi \rho_{\text{P}} - 512}{8 (\pi \rho_{\text{P}} - 4)^3}. 
\label{eq_Dc} 
\end{equation*}
Steady-state solutions of Eqs.~(\ref{eq_active}) and (\ref{eq_passive}) are obtained by transforming the equations into polar coordinates, leveraging system symmetry, and applying approximate boundary conditions. These solutions provide the radial particle number densities and the four components of the state-switching time, shown as fitting lines in Fig.~\ref{fig_density}(b-d) and Fig.~\ref{fig_duration}(a). Details of the calculation are provided in the Supplemental Material.



After the transition, $\langle T_{\text{P}}^{\text{N}} \rangle$ and $\langle T_{\text{P}}^{\text{L}} \rangle$ show slight increases with $N$, but remain relatively small [Fig.~\ref{fig_duration}(a)]. This behavior can be attributed to the increased number of active particle layers, which exert greater stress on the passive layers, making it more difficult for passive particles to exit the loss region and pass through the neutral region.
In contrast, $\langle T_{\text{A}}^{\text{N}} \rangle$ increases significantly with $N$ and becomes the dominant contributor to $\langle T \rangle$. This is due to two factors: first, the slight increase in $\langle T_{\text{P}}^{\text{L}} \rangle$ leads to a lower exchange rate between active and passive particles, as the number of particles in the loss region remains roughly constant—each time a passive particle exits the loss region, an active particle enters. Second, a larger $N$ results in the formation of more active particle layers outside the loss region, increasing the average queuing time. Additionally, $\langle T_{\text{A}}^{\text{G}} \rangle$ shows a slight increase after the transition, as active particles spend more time in the neutral region, increasing the likelihood of winding around and passing through the gain region multiple times before entering the loss region [see typical trajectory in Fig.~\ref{fig_duration}(e)].

Interestingly, we observe spatial clustering of particles transitioning from the passive to active states. In Fig.~\ref{fig_clustering}, passive particles entering the gain region within $2\tau$ are color-coded using a gradient scale. Before the transition, Figs.~\ref{fig_clustering}(a-b) show that similarly colored passive particles are uniformly distributed. After the transition, Fig.~\ref{fig_clustering}(c) reveals that, although some particles still transition individually, many form spatial clusters, with particles transitioning to the active state sequentially. See Supplemental Movies for Figs.~\ref{fig_clustering}(a-c) and Supplemental Material for additional snapshots at other $L_2$ values.
We attribute this clustering to the system's inhomogeneity and fluctuations. Active particles at the edges of passive layers exert uneven stress on passive particles. Defects and voids in these layers can cause local stress imbalances, which can facilitate the movement of passive particles. Once passive particles exit some of the active layers, the number of remaining active layers that exert obstructive stress decreases, reducing resistance and facilitating further movement. Moreover, the exit of passive particles lowers pressure in vacated regions, and passive particles in adjacent regions are then pushed toward these low-stress areas, encouraging clustering behavior. This aligns with the reduced randomness and fewer self-intersections observed in individual particle trajectories after the transition [Fig.~\ref{fig_duration}(e)].

In conclusion, we investigate a minimal model of self-propelled Brownian particles undergoing cyclic transitions between active and passive states across distinct regions. We identify a structural transition that marks a threshold particle number. Below this threshold, increasing the particle number reduces the state-switching time, while above the threshold, further increases in particle number lead to longer switching times. Additionally, above the threshold, we observe the emergence of spatial clustering, where particles transitioning from passive to active states tend to do so in close proximity. These findings provide insights into the collective dynamics of cyclic processes between active and passive states in active matter systems.


\begin{acknowledgments}

This work was supported by the National Natural Science Foundation of China (Grant No.~12274330). D.W. acknowledges the ``Xiaomi Young Scholar Program" at Wuhan University. 

\end{acknowledgments}

\bibliography{ActiveMatter}

\end{document}


\title{\large Supplemental Materials for:\\
Self-propelled particles undergoing cyclic transitions}
\author{Ye Zhang}
\affiliation{Key Laboratory of Artificial Micro- and Nano-structures of Ministry of Education and School of Physics and Technology, Wuhan University, Wuhan 430072, China}
\author{Duanduan Wan}
\email[E-mail: ]{ddwan@whu.edu.cn}
\affiliation{Key Laboratory of Artificial Micro- and Nano-structures of Ministry of Education and School of Physics and Technology, Wuhan University, Wuhan 430072, China}
\maketitle

\setcounter{equation}{0}
\setcounter{figure}{0}
\setcounter{table}{0}
\setcounter{page}{1}
\makeatletter
\renewcommand{\theequation}{S\arabic{equation}}
\renewcommand{\thefigure}{S\arabic{figure}}
\renewcommand{\thetable}{S\arabic{table}}
\textbf{This PDF includes:}


\begin{itemize}
    \item Captions for Movies S1 to S3
    \item Snapshots of the steady-state configurations
    \item Calculation of particle number densities in the radial direction
    \item Derivation of the analytical model and its solution
    \item Calculation of the state-switching time and its components
\end{itemize}

\vspace{1em} 

\textbf{Other supplemental materials for this manuscript include the following:}

\begin{itemize}
    \item Movies S1 to S3 (.mp4)
\end{itemize}

\newpage
\section{I. Captions for Movies S1 to S3}
Movies S1 to S3 show the transition of particles from passive to active states for $L_2 = 5$ and $N = 1000$, $N = 2000$, and $N = 3000$, respectively. Active particles are shown in pink, while passive particles that will transition to active states within the next $2\tau$ are color-coded according to a gradient: green particles will activate soon, and blue particles will activate closer to $2\tau$. Other passive particles are depicted in grey. The parameters and color-coding scheme match those in Fig.~4 of the main text. The movie files are as follows:
\begin{itemize}
    \item Movie\_S1\_N\_1000\_L2\_5.mp4
    \item Movie\_S2\_N\_2000\_L2\_5.mp4
    \item Movie\_S3\_N\_3000\_L2\_5.mp4
\end{itemize}

\section{II. Snapshots of the steady-state configurations}

\begin{figure}[h]
    \centering
    \includegraphics[width=0.45\linewidth]{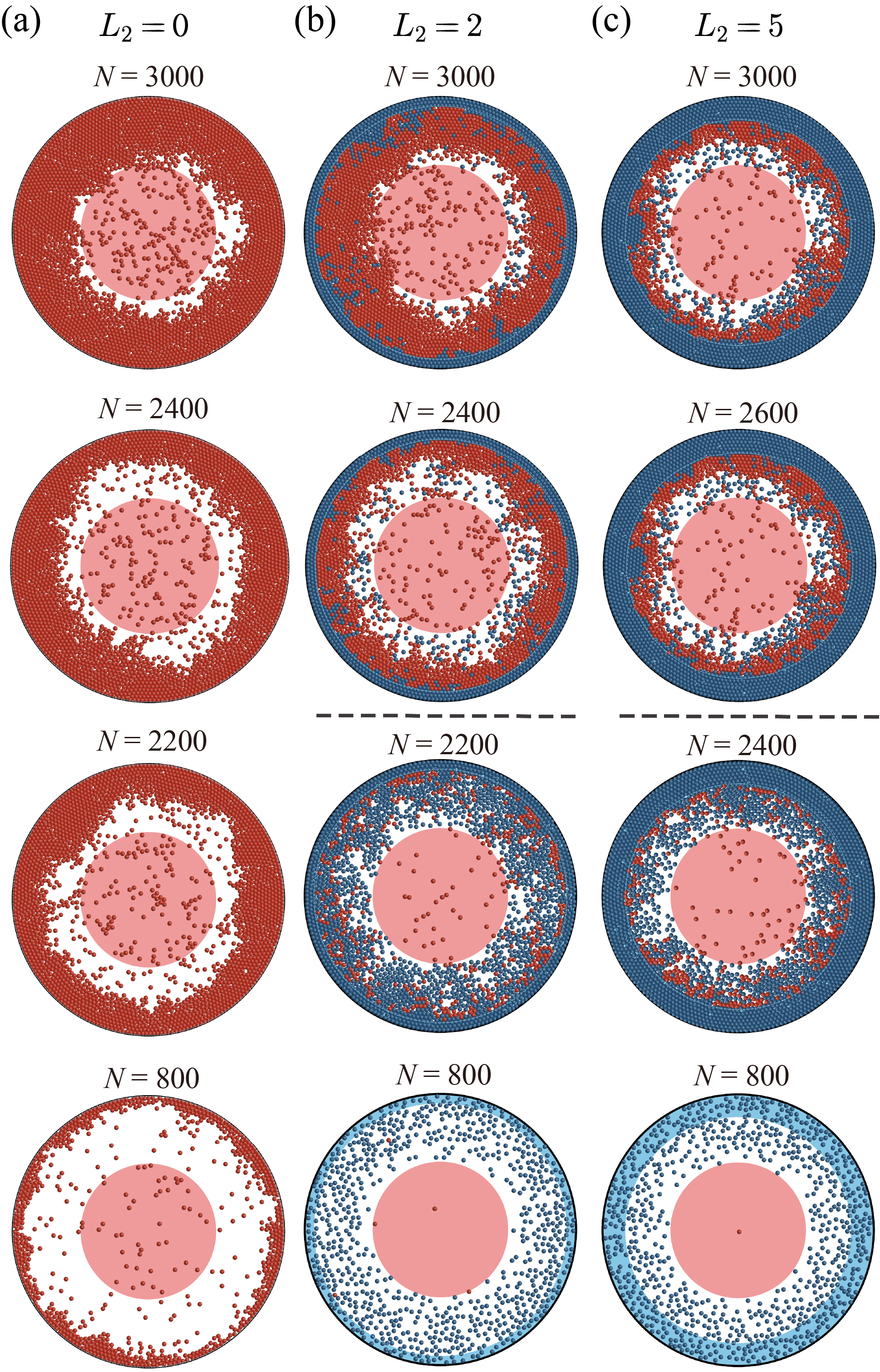}
    \caption{Snapshots of the steady-state configurations at selected values of $L_2$ and $N$. The grey dashed line marks the transition positions shown in Fig.~1(b) of the main text. }
    \label{fig_snapshot_full}
\end{figure}

Figure \ref{fig_snapshot_full}(a) shows snapshots of the steady-state configurations for $L_2=0$ at selected values of $N$. In this case, with no loss area, all particles eventually become active. As $N$ increases, additional layers of particles form at the boundary of the system. Figures \ref{fig_snapshot_full}(b) and (c) show snapshots of the system for $L_2=2$ and $L_2=5$, respectively, at values of $N$ that are far below, just below, just above, and far above the transition threshold.
For $L_2=2$, at small $N$ (e.g., $N=800$), most particles remain in the passive state, sparsely distributed in the loss and neutral regions, with only a small fraction of active particles in the gain and neutral areas (approximately $0.3\%$ active). This trend persists as $N$ increases, approaching the transition. Just below the transition, at $N=2200$, the loss region is fully occupied by passive particles, while particle accumulation increases in the neutral area near the loss region. The fraction of active particles also rises. After the transition, at $N=2400$, there is a sharp increase in the number of active particles, which accumulate near the boundary of the loss region. For even larger $N$ values, such as $N=3000$, additional layers of active particles form. A similar behavior is observed for $L_2=5$, though here, more layers of active particles emerge within the loss region as $N$ nears or exceeds the transition value.

\begin{figure}[h]
    \centering
    \includegraphics[width=0.55\linewidth]{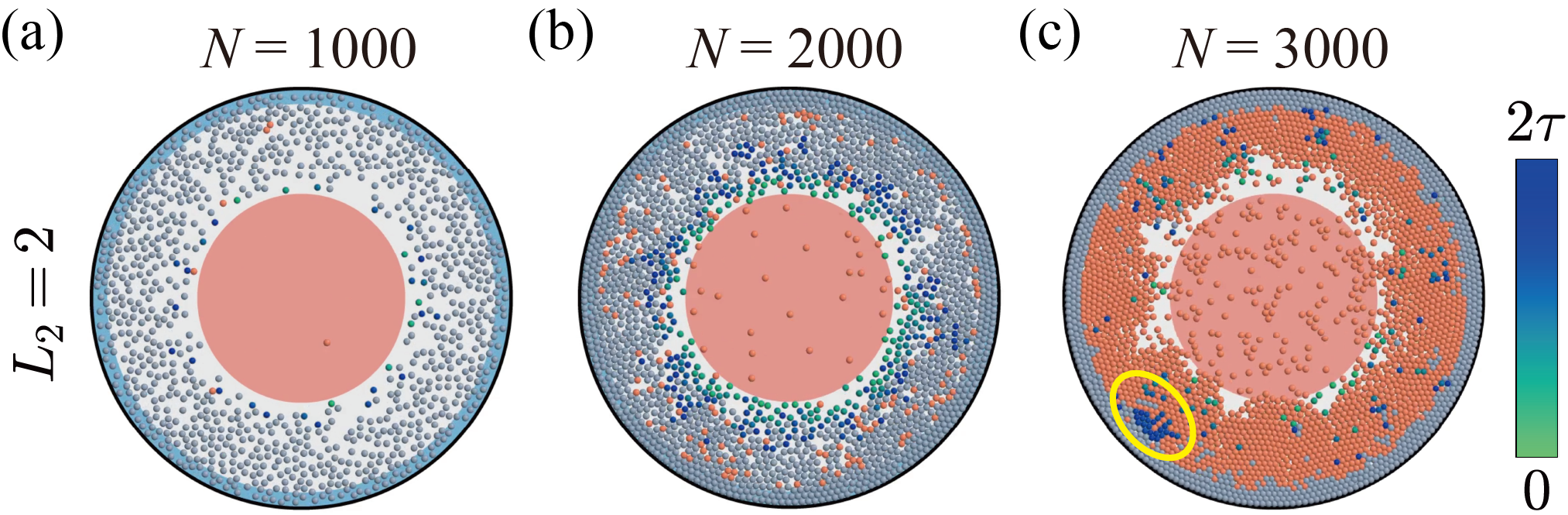}
    \caption{Sequence of particles transitioning from passive to active states for $L_2 = 2$, with the color-coding scheme matching that in Fig.~4 of the main text.}
    \label{fig_cluster_SM}
\end{figure}

Figure \ref{fig_cluster_SM} illustrates the sequence of particles transitioning from passive to active states for $L_2 = 2$, with the color-coding scheme consistent with that in Fig.~4 of the main text.

\section{III. Calculation of particle number densities in the radial direction}
We define the radial particle number density, $\rho(r)$, which quantifies the number of particles located between distances $r$ and $r + \Delta r$ from the center of a circular region. Assuming the particles are disks with diameter $\sigma$, $\rho(r)$ is given by
\begin{equation} 
\rho(r) = \frac{4}{\pi \sigma^2} \frac{A(r + \Delta r) - A(r)}{\pi \left[ (r + \Delta r)^2 - r^2 \right]}, \end{equation}
where $A(r)$ denotes the area within the circular region of radius $r$ that is covered by particles. The quantity $\rho(r)$ quantifies the particle coverage in the annular region between $r$ and $r + \Delta r$, normalized by the area of a single particle, $a_0 = \pi \sigma^2 / 4$, and divided by the area of the annulus. We take $\Delta r = 0.1$ for numerical calculations. The particle number densities in the passive and active states are denoted as $\rho_\text{P}(r)$ and $\rho_\text{A}(r)$, respectively.
Snapshots are collected every $0.01 \tau$ after the system reaches steady state. To ensure adequate sampling, the number of snapshots is adjusted based on the total particle count $N$, with the total number of snapshots given by $1 \times 10^7 / N$ for a system containing $N$ particles.







\section{III. Derivation of the analytical model and its solution}

We denote $s = \text{A}, \text{P}$ to indicate particles in the active and passive states, respectively. The probability density function $\chi_i(x, y, \theta, s, t)$ represents the probability density of particle $i$ being at position $(x, y)$, oriented at angle $\theta$, and in state $s$ at time $t$. Normalization is ensured by the condition  
\begin{equation}
\sum_{s} \iint \text{d}x \text{d} y \int_{0}^{2\pi} d\theta\, \chi_i(x, y, \theta, s, t) = 1.
\label{eq_normalization}
\end{equation}
Let $\psi_{\text{A},i}$ and $\psi_{\text{P},i}$ denote the probability density for particle $i$ in the active and passive states, respectively, i.e., $\psi_{\text{A},i }(x,y,\theta,t)=\chi_i(x,y,\theta,s=A,t)$ and $\psi_{\text{P},i}(x,y,\theta,t)=\chi_i(x,y,\theta,s=P,t)$. $\psi_{\text{A},i}$ and $\psi_{\text{P},i}$ relate to the particle number density functions as $\rho_{{\text A}}(x,y,t)=\sum_{i} \rho_{{\text A}, i}(x,y,t)=\sum_{i}\int_{0}^{2\pi}\psi_{\text{A,}i}\left(x,y,\theta,t\right){\text d}\theta$ and $\rho_{{\text P}}(x,y,t)=\sum_{i} \rho_{{\text P}, i}(x,y,t)= \sum_{i}\int_{0}^{2\pi}\psi_{\text{P,}i}\left(x,y,\theta,t\right){\text d}\theta$, respectively.

\begin{figure}
\centering
\includegraphics[width=0.5\linewidth]{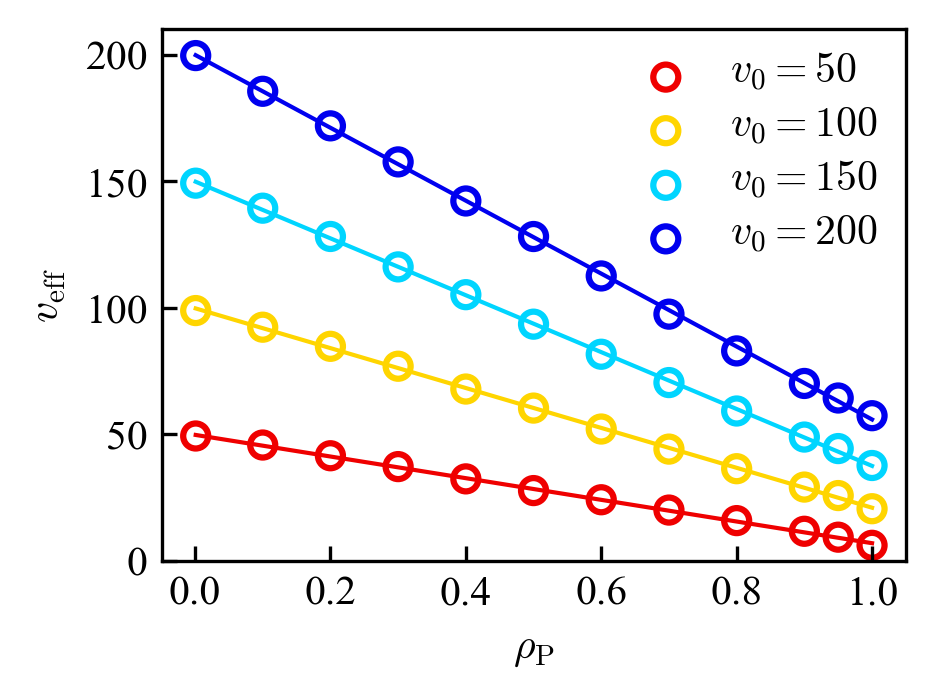}
\caption{The effective self-propulsion speed, $v_\text{eff}$, of an active particle immersed in a system of 1023 passive particles. $v_\text{eff}$ is measured as defined in Eq.~(\ref{Eq:veff}). The lines correspond to fits of the form $v_\text{eff} = (1 - c \rho_\text{P}) v_0$. For $v_0 = 50$, $100$, $150$, and $200$, the corresponding values of the slope $c$ are $0.86$, $0.79$, $0.75$, and $0.72$, respectively. }
\label{fig_veff}
\end{figure}

We develop a model to analyze scenarios below the transition, where only a small fraction of the total particle population remains active in the steady state, as observed in simulations. Under these conditions, the diffusion equation for the active-state probability density, $\psi_{\text{A},i}$, is given by:
\begin{equation}
    \frac{\partial \psi_{\text{A},i}(x, y, \theta, t)}{\partial t} = \nabla \cdot \left( - v_{\text{eff}} \mathbf{n}_i \, \psi_{\text{A},i} \right) + D_{\text{r}} \frac{\partial^2 \psi_{\text{A},i}}{\partial \theta^2}.
\label{eq_active}
\end{equation}
Here, $v_{\text{eff}}$ is the effective self-propelled velocity of an active particle. This effective velocity deviates from its original value $v_0 = D_\text{t}\beta f_0$ due to collisions with passive particles.
We determine the effective velocity, $v_{\text{eff}}$, by measuring the instantaneous velocity projected onto the self-propulsion direction:
\begin{equation}
v_{\text{eff}} = \langle \dot{\mathbf{r}}_i \cdot \mathbf{n}_i \rangle = \langle (\mathbf{d}_i / \text{d} t) \cdot \mathbf{n}_i \rangle.
\label{Eq:veff}
\end{equation}
Here $\mathbf{d}_i$ denotes the displacement of particle $i$ over a time step $\text{d}t$. We set $\text{d} t = 10^{-6}\tau$, consistent with the value used in the MD simulations discussed in the main text, and the average is taken over a time span of $t = 10\tau$. Figure \ref{fig_veff} illustrates the effective velocity $v_{\text{eff}}$ as a function of the passive particle density $\rho_{\text{P}}$. The data reveal a linear decay of $v_{\text{eff}}$ with increasing $\rho_{\text{P}}$, with the slope depending on $v_0$. This trend is consistent with previous reports \cite{Stenhammar_2013_Continuum, Stenhammar_2015_ActivityInduced}. Fitting the data to the linear form $v_{\text{eff}} = (1 - c \rho_{\text{P}}) v_0$, with $v_0 = 150$, yields a slope parameter $c = 0.75$.

Next, we describe the diffusion equation for the passive-state probability density $\psi_{\text{P},i}$ as
\begin{equation}
    \frac{\partial \psi_{\text{P},i}(x, y, \theta, t)}{\partial t} = \nabla \cdot \left( D_{\text{c}}(\rho_{\text{P}}) \nabla \psi_{\text{P},i} - \sum_{j} \int_{0}^{2\pi} \psi_{\text{A}, j}(x, y, \theta, t) c v_0 \mathbf{n}_j \, d\theta \, \psi_{\text{P},i} \right) + D_{\text{r}} \frac{\partial^2 \psi_{\text{P},i}}{\partial \theta^2}.
\label{eq_passive}
\end{equation}
The first term on the right-hand side represents the collective diffusion term, where \( D_{\text{c}}(\rho_{\text{P}}) \) is the collective diffusion coefficient, which depends on the passive particle density. The collective diffusion coefficient \( D_{\text{c}} \) is related to \( D_\text{t}\) by the expression \( D_{\text{c}} = D_t/S(0) \), where \( S(0) \) is the static structure factor \( S(k) \) in the limit \( k \to 0 \) \cite{Hess_1983_Generalized, Lahtinen_2001_Diffusion}. Additionally, \( S(0) \) can be related to the isothermal compressibility \( \kappa_{\text{T}} \) as $S(0) = \kappa_{\text{T}} \rho/\beta $ \cite{Hess_1983_Generalized, Lahtinen_2001_Diffusion}. We compute \( \kappa_{\text{T}} \) using the equation of state for hard disks from Ref.~\cite{Douglas_1975_ASimple}. Finally, the collective diffusion coefficient is given by:
\begin{equation}
    D_{\text{c}}(\rho_{\text{P}}) = \frac{D_{\text{t}} \beta}{\rho_{\text{P}} \kappa_{\text{T}}} = \frac{\pi^3 \rho_{\text{P}}^3 - 12 \pi^2 \rho_{\text{P}}^2 - 128 \pi \rho_{\text{P}} - 512}{8 (\pi \rho_{\text{P}} - 4)^3}.
\label{eq_Dc}
\end{equation}
The second term on the right-hand side of Eq.~(\ref{eq_passive}) accounts for the effective force exerted by active particles on a passive particle. Specifically, based on the expression for $v_{\text{eff}}$, we consider that an active particle $j$ experiences an obstructive force of $c \rho_{\text{P}} v_0 / D_\text{t}\beta$ in the direction opposite to its self-propulsion direction, due to collisions with passive particles. Therefore, each passive particle receives a reaction force of $c v_0 / D_\text{t}\beta$ in the direction of the self-propulsion of the active particle $j$. For systems where active particles make up only a small fraction of the total population in the steady state (i.e., $\rho_{\text{A}}$ is small), we neglect the cross-diffusion term, which would otherwise account for the effect of $\nabla \rho_{\text{A}}$ on the diffusion of passive particles \cite{Bruna_2012_Diffusion}.

\begin{figure}
\centering
\includegraphics[width=0.3\linewidth]{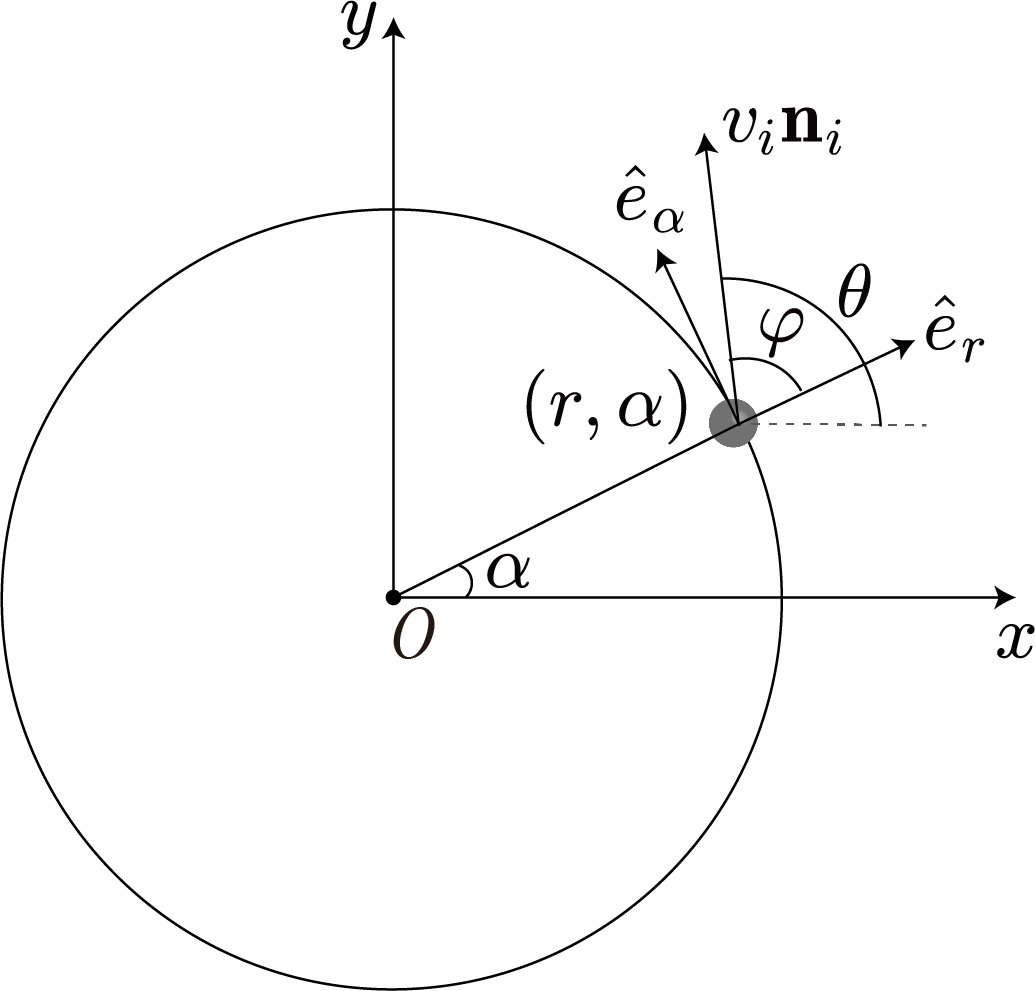}
\caption{Relevant angles considered in this study within the polar coordinate system.}
\label{fig_polar_coordinate}
\end{figure}

\begin{figure}
    \centering
    \includegraphics[width=0.8\linewidth]{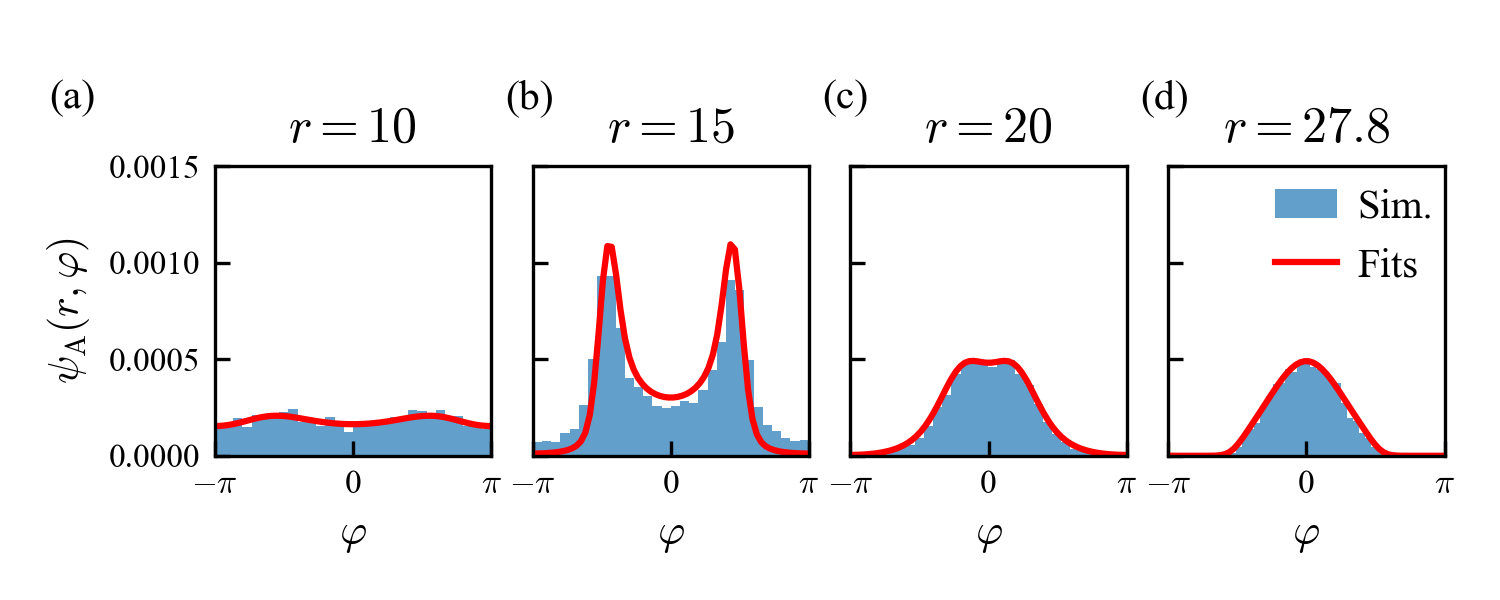}
    \caption{Distribution of $\psi_\text{A}(r, \varphi)$ with respect to $\varphi$ at selected values of $r$  for $N=800$ and $L_2=2$. The blue histograms represent simulation results, with particles within a radial range of $r \pm 0.2$ are considered. The red lines correspond to model predictions.}
    \label{fig_psiAphi}
\end{figure}

To obtain the constant solutions of Eq.~(\ref{eq_active}) and Eq.~(\ref{eq_passive}) in the steady state, we perform a coordinate transformation from Cartesian to polar coordinates. Specifically, we transform $\psi_{\text{A},i}(x, y, \theta, t)$ and $\psi_{\text{P},i}(x, y, \theta, t)$ into $\psi_{\text{A},i}(r, \alpha, \varphi, t)$ and $\psi_{\text{P},i}(r, \alpha, \varphi, t)$, respectively. Here, $r$ and $\alpha$ denote the radial and angular coordinates in the polar system, while $\varphi = \theta - \alpha$ represents the relative angle between the self-propulsion direction and the polar angle. The relationships between these angles are illustrated in Fig.~\ref{fig_polar_coordinate}. In polar coordinates, the gradient operator is expressed as:
\begin{equation}
\nabla =\hat{e}_{r}\frac{\partial}{\partial r}+\hat{e}_{\alpha}\frac{1}{r}\frac{\partial}{\partial\alpha}.
\label{eq_gradient_operator}
\end{equation}
Substituting this into Eq.~(\ref{eq_active}) and noting that $\psi_{\text{A},i}(r, \alpha, \varphi, t) = \psi_{\text{A},i}(r, \varphi, t)$, i.e., $\psi_{\text{A},i}$ is independent of $\alpha$ owing to the circular symmetry of the system, we obtain:
\begin{equation}
\frac{\partial \psi_{\text{A},i}(r, \varphi, t) }{\partial t}= -v_{0}\cos\varphi\frac{\partial}{\partial r}\psi_{\text{A},i}(1-c\rho_{\text{P}})+\frac{1}{r}v_{0}\sin\text{\ensuremath{\varphi}}\frac{\partial}{\partial\varphi}\psi_{\text{A},i}(1-c\rho_{\text{P}}) + D_{r}\frac{\partial\psi_{A,i}}{\partial\varphi^{2}}.
\end{equation}
Similarly, for passive particles, substituting Eq.~(\ref{eq_gradient_operator}) into Eq.~(\ref{eq_passive}) and noting $v_{0}\mathbf{n}_j = v_{0}\cos\varphi\,\hat{e}_{r} + v_{0}\sin\varphi\,\hat{e}_{\alpha}$ in Eq.~(\ref{eq_passive}), we arrive at:
\begin{equation}
\frac{\partial\psi_{\text{P},i}(r,\varphi,t)}{\partial t}
=  (\frac{\partial}{\partial r} + \frac{1}{r}) \left(D_{\text{c}}(\rho_{P})\frac{\partial}{\partial r}\psi_{\text{P},i}-c\psi_{\text{P},i} \sum_{j} \int_{0}^{2\pi}\psi_{\text{A},j}v_{0}\cos\varphi d\varphi\right)+D_{\text{r}}\frac{\partial^{2}\psi_{\text{P},i}}{\partial\varphi^{2}}.
\end{equation}
In the steady state, we consider $m$ particles entering the gain region from the neutral region per unit time. Due to the conservation of the total number of active particles in the gain region, there are also $m$ particles leaving this region in the same time interval. The overall effect is that the neutral region loses $m$ passive particles while gaining $m$ active particles. Consequently, the conservation of active and passive particles in the neutral region implies a corresponding exchange of $m$ particles at the boundary of the loss region.
For active particles, the $m$ particles entering the gain region from the neutral region per unit time act as a source at the gain boundary. This leads to the governing equation with the boundary conditions as:

\begin{equation}
\begin{cases}
\begin{aligned}0= &- v_{0}\cos\varphi\frac{\partial\psi_{\text{A},i}(1-c\rho_{\text{P}})}{\partial r}+\frac{1}{r} v_{0}\sin\varphi\frac{\partial\psi_{\text{A},i}\left(1-c\rho_{\text{P}}\right)}{\partial\varphi} 
\\&+ D_{\text{r}}\frac{\partial\psi_{\text{A},i}}{\partial\varphi^{2}}+\frac{m/N}{(2\pi)^{2}L_{1}}\delta\left(L_{1}\right),\\[8pt]
\end{aligned}
 & 0<r<R-L_{2},0<\varphi<2\pi,\\[10pt]
\psi_{\text{A},i}(r,\varphi,t)\Big|_{r=R-L_{2}}=0, & 0<\varphi<2\pi, \\[10pt]
\psi_{\text{A},i}(r,\varphi,t)\Big|_{\varphi=2\pi}=\psi_{\text{A},i}(r,\varphi,t)\Big|_{\varphi=0}, & 0<r<R-L_{2},\\[10pt]
\frac{\partial\psi_{\text{A},i}(r,\varphi,t)}{\partial\varphi}\Big|_{\varphi=2\pi}=\frac{\partial\psi_{\text{A},i}(r,\varphi,t)}{\partial\varphi}\Big|_{\varphi=0}, & 0<r<R-L_2.\\
\end{cases}
\label{eq_active_bcs}
\end{equation}
The choice of the source term satisfies the normalization condition: 
\begin{equation}
\int_{0}^{R}r\text{d}r\int_{0}^{2\pi}\text{d}\alpha\int_{0}^{2\pi}\text{d}\varphi\frac{m/N}{(2\pi)^{2}L_{1}}\delta\left(L_{1}\right)=m/N.    
\end{equation}
Similarly, for passive particles, the governing equation can be simplified to:
\begin{equation}
\begin{aligned}
\frac{\partial\psi_{\text{P},i}(r,\varphi,t)}{\partial t} = & \left(\frac{\partial}{\partial r} + \frac{1}{r}\right)\left(D_{\text{c}}(\rho_{P})\frac{\partial}{\partial r}\psi_{\text{P},i} 
    - c\psi_{\text{P},i}\sum_{j}\int_{0}^{2\pi}\psi_{\text{A},j}v_{0}\cos\varphi \, d\varphi\right) \\
    & + D_{\text{r}}\frac{\partial^{2}\psi_{\text{P},i}}{\partial\varphi^{2}} 
    + \frac{f(\varphi)m/N}{(2\pi)^{2}(R-L_{2})}\delta\left(r-(R-L_{2})\right).
\label{eq_passive_phi}
\end{aligned}
\end{equation}
Here the function $f(\varphi)$ in the source term represents the non-uniform distribution of particles transitioning from active to passive at the loss boundary with respect to $\varphi$, where $\int_{0}^{2\pi} f(\varphi)\text{d}\varphi = 2\pi$. The source term satisfies the normalization condition:
\begin{equation}
\begin{aligned}
\int_{0}^{R}r\text{d}r\int_{0}^{2\pi}\text{d}\alpha\int_{0}^{2\pi}\text{d}\varphi\frac{f(\varphi)m/N}{(2\pi)^{2}(R-L_{2})}\delta\left(r-(R-L_{2})\right)= & \int_{0}^{R}r\text{d}r\int_{0}^{2\pi}\text{d}\alpha\frac{m/N}{2\pi(R-L_{2})}\delta\left(r-(R-L_{2})\right)\\
= & m/N.
\end{aligned}
\end{equation}
To further simplify the calculation, we integrate Eq.~(\ref{eq_passive_phi}) with respect to $\varphi$ from $0$ to $2\pi$, summing over all passive particles. In the steady state, this leads to:
\begin{equation}
\begin{cases}
\begin{aligned}0= & \left(\frac{\partial}{\partial r} +\frac{1}{r}\right)
\left(D_{\mathrm{c}}\left(\rho_{\mathrm{P}}\right)\frac{\partial}{\partial r}\rho_{\mathrm{P}} 
- c \rho_{\mathrm{P}} \sum_{j} \int_{0}^{2\pi} \psi_{\mathrm{A},j} v_{0} \cos\varphi \, d\varphi\right)
\\& +\frac{m}{2\pi\left(R - L_{2}\right)} \delta\left(r - \left(R - L_{2}\right)\right) ,\\[8pt]
\end{aligned}
 & L_{1} < r < R - \sigma/2,\\[10pt]
{\displaystyle \rho_{\mathrm{P}}(r,t)\big|_{r=L_{1}} = 0,} \\[10pt]
{\displaystyle \frac{\partial \rho_{\mathrm{P}}(r,t)}{\partial r} \big|_{r = R - \sigma/2} = 0.}
\end{cases}
\label{eq_passive_rho}
\end{equation}
We numerically solve Eqs.~(\ref{eq_active_bcs}) and (\ref{eq_passive_rho}), with $D_{\text{c}}(\rho_{\text{P}})$ defined by Eq.~(\ref{eq_Dc}). We calculate $\psi_\text{A}(r, \varphi)$ as the sum of individual contributions: $\psi_\text{A}(r, \varphi) = \sum_i \psi_{\text{A},i}(r, \varphi) = N \psi_{\text{A},i}(r, \varphi)$.  Figure~\ref{fig_psiAphi} presents an example with $N = 800$ and $L_2 = 2$, showing $\psi_\text{A}(r, \varphi)$ at selected values of $r$. The solution for $\rho_{\text{P}}(r)$ provides the fitting lines shown in Fig.~2 of the main text.

\subsection{IV. Calculation of the state-switching time and its components}

The mean number of particles in different states and regions—specifically, the mean number of particles in the passive state within the loss region, $\langle N_{\text{P}}^{\text{L}}\rangle$, in the passive state within the neutral region, $\langle N_{\text{P}}^{\text{N}}\rangle$, in the active state within the neutral region, $\langle N_{\text{A}}^{\text{N}}\rangle$, and in the active state within the gain region, $\langle N_{\text{A}}^{\text{G}}\rangle$—can be calculated using the particle density functions as follows:
\begin{equation}
\langle N_{\text{P}}^{\text{L}} \rangle = 
N \int_{R-L_{2}}^{R-\sigma/2} \text{d}r \int_{0}^{2\pi} \text{d}\alpha \int_{0}^{2\pi} \text{d}\varphi \, \psi_{\text{P},i}(r, \varphi)
\end{equation}

\begin{equation}
\langle N_{\text{P}}^{\text{N}} \rangle = 
N \int_{L_{1}}^{R-L_{2}} \text{d}r \int_{0}^{2\pi} \text{d}\alpha \int_{0}^{2\pi} \text{d}\varphi \, \psi_{\text{P},i}(r, \varphi)
\end{equation}

\begin{equation}
\langle N_{\text{A}}^{\text{N}} \rangle = 
N \int_{L_{1}}^{R-L_{2}} \text{d}r \int_{0}^{2\pi} \text{d}\alpha \int_{0}^{2\pi} \text{d}\varphi \, \psi_{\text{A},i}(r, \varphi)
\end{equation}

\begin{equation}
\langle N_{\text{A}}^{\text{G}} \rangle = 
N \int_{0}^{L_{1}} \text{d}r \int_{0}^{2\pi} \text{d}\alpha \int_{0}^{2\pi} \text{d}\varphi \, \psi_{\text{A},i}(r, \varphi)
\end{equation}
Additionally, the parameter $m$, representing the number of particles entering the gain region from the neutral region per unit time [see Eq.~(\ref{eq_active_bcs})], allows us to calculate the average state-switching time, $\langle T \rangle$, using the formula:
\begin{equation}
\langle T \rangle= \frac{N}{m}.
\end{equation}
The contributions of various regions and states to the state-switching time, represented as $\langle T_{\text{P}}^{\text{L}} \rangle$, $\langle T_{\text{P}}^{\text{N}} \rangle$, $\langle T_{\text{A}}^{\text{N}} \rangle$, and $\langle T_{\text{A}}^{\text{G}} \rangle$, can be calculated using the following relations:  
\begin{align}
\langle T_{\text{P}}^{\text{L}} \rangle &= \frac{\langle N_{\text{P}}^{\text{L}} \rangle}{N} \langle T \rangle, \\  
\langle T_{\text{P}}^{\text{N}} \rangle &= \frac{\langle N_{\text{P}}^{\text{N}} \rangle}{N} \langle T \rangle, \\  
\langle T_{\text{A}}^{\text{N}} \rangle &= \frac{\langle N_{\text{A}}^{\text{N}} \rangle}{N} \langle T \rangle, \\  
\langle T_{\text{A}}^{\text{G}} \rangle &= \frac{\langle N_{\text{A}}^{\text{G}} \rangle}{N} \langle T \rangle.  
\end{align}  
These provide the fitting lines shown in Fig.~3(a) of the main text.

\bibliography{ActiveMatter.bib}